\documentclass[12pt]{JHEP3}

\usepackage{amsmath,amssymb,graphicx,epsfig,subeqnarray}%,showkeys}

\newcommand{\be}{\begin{equation}}
\newcommand{\ee}{\end{equation}}
\newcommand{\ba}{\begin{eqnarray}}
\newcommand{\ea}{\end{eqnarray}}
\newcommand{\bea}{\begin{array}}
\newcommand{\eea}{\end{array}}

\newcommand{\Tr}{{\rm Tr}}
\newcommand{\BW}{{\bar{W}}}

\def\Tr{{\rm Tr}}

\def\I_M{{I_{\scriptscriptstyle M\times M}}}

\preprint{KIAS-P007075}

\title{New BPS Configurations of  BMN Matrix
Theory}

\author{ Jens Hoppe$^{*a}$
and Ki-Myeong Lee$^{\dag b}$   \\
\\$^*$ Department of Mathematics,
Royal Institute of Technology \\ \hspace{0.2cm}  SE-10044 Stockholm, Sweden \\
$^\dag$ School of Physics,
  Korea Institute for Advanced Study \\
  \hspace{0.2cm}  Seoul  130-722,
Korea\\ $~$ \\ $^a$ E-mail:  hoppe@kth.se  \\ $^b$ E-mail:
klee@kias.re.kr}

\abstract{ We explore the 1/2 BPS configurations in BMN matrix
theory with $SO(3)$ angular momentum of $SO(3)\times SO(6)$
symmetry. The fluctuation analysis of the BPS configurations near
the abelian solutions and also the fuzzy two sphere vacua reveals
how nonabelian BPS configurations emerge. Especially the
irreducible nonabelian configurations seem to have the maximal
angular momentum of order $N^3$, beyond which they collapse to
abelian ones. We also find some new BPS configurations
explicitly.}

\begin{document}

\section{Introduction}

BMN matrix theory with $U(N)$ gauge group has been proposed as the
DLCQ limit of M theory on the 11-dim plane wave background with
the maximal supersymmetry~\cite{Berenstein:2002jq}. The action of
this model can be also obtained from the matrix regularization of
the membrane action in the pp-wave background, or the quantum
mechanics of D0 branes on the background of the 11-d pp-wave
compactified to 10-dim ~\cite{Dasgupta:2002hx}, or by dimensional
reduction of 4-dim susy Yang-Mills theory on $R\times
S^3$~\cite{Kim:2003rza}. Each vacuum of the BMN matrix theory is
characterized by a partition of $N$ and describes concentric fuzzy
2-spheres, which can be interpreted as giant gravitons in the
pp-wave background.

The matrix theory has $SO(3)\times SO(6)$ symmetry and these fuzzy
2-sphere vacua can have also $SO(3)$ angular momentum, breaking
the supersymmetry partially. While  the BPS configurations with
$SO(6)$ angular momentum are trivial ones, the BPS configurations
with $SO(3)$ angular momentum are highly nontrivial. Only a few
exact BPS configurations have been found for finite $N$
~\cite{Bak:2002rq,Mikhailov:2002wx,Bak:2005jh}. In the infinite
$N$ ( continuum ) limit  the general BPS configurations
  have been found to be Riemann surface with arbitrary number of
  genus and spikes~\cite{Bak:2005jh}.  The BPS configurations for
finite $N$, which are expected to be a special class of fuzzy
Riemann surfaces,
  have not been understood in general.  ( see however
Ref.~\cite{Arnlind:2006ux}. )

In this work, we investigate  these 1/2 BPS configurations
carrying $SO(3)$ angular momentum for finite $N$ in BMN matrix
theory. A detailed fluctuation analysis of the BPS configurations
near  the abelian BPS configurations and also near the nonabelian
ground states leads to new insight on  how nonabelian BPS
configurations emerge. Based on this analysis, we make several
observations about these 1/2 BPS configurations of the BMN matrix
theory. One is about the  maximum value of the $SO(3)$ angular
momentum for any irreducible nonabelian configuration, which is
defined as one which cannot be expressed as a sum of commuting
configurations ). We also find a   new explicit class of BPS
configurations for higher $N$.

The simplest BPS configurations with $SO(3)$ angular momentum are
  abelian solutions which are present even in $U(1)$ theory. In the
$U(N)$ theory, the abelian BPS configuration would be made out of
the time-dependent field configurations which are all
simultaneously diagonaliable. The next simplest ones are the
ellipsoidal solutions  which are time-dependent configurations
built on vacuum fuzzy spheres ~\cite{Bak:2002rq}. These fuzzy
ellipsoidal solutions become abelian when the angular momentum,
say  $J_3$, exceeds a critical value of order $N^3$. (A similar
phenomena concerning the maximal angular momentum of nonabelian
angular momentum has been observed recently, in a somewhat
different context~\cite{Iengo:2007cp}.)

For small finite $N$, some toroidal BPS configurations have
previously been found
explicitly~\cite{Mikhailov:2002wx,Bak:2005jh}. All 1/2 BPS
solutions with finite $SO(3)$ angular momentum have been also
found in the continuum limit with infinite $N$~\cite{Bak:2005jh}.
These continuum BPS configurations consist of all possible genus
surfaces maybe with some spikes.  Our BPS equations do not seem
integrable for finite $N$.
  All BPS solutions for $N=2$ are known but we think that not all the
solutions with $N=3$ are known.  We do a detailled analysis of the
BPS configurations lying near  abelian solutions or vacuum fuzzy
spheres. This analysis shows that nonabelian BPS configurations
emerge from the known solutions in a very specific way.

BMN matrix theory and BPS states have been analyzed in detail from
the superalgebra analysis and the perturbative approach of the BPS
states by expanding the theory around each vacuum in the large
$\mu$ limit~\cite{Dasgupta:2002hx,Dasgupta:2002ru,Kim:2002if}.
  The infinite $\mu$ limit is a free theory and
the interaction is of order $1/\mu$. This consideration does not
directly involve  the classical BPS configurations, which are
nonperturbative. The protected spectrum of the matrix theory in
the large $N$ limit is also related to  the linear fluctuation
spectrum of the spherical M5-brane~ \cite{Maldacena:2002rb}. Our
BPS configurations in the large $N $ limit should also play a role
in this context.

There are several degenerate vacua in the theory which are
separated by an energy barrier and the tunnelling between these
vacua has been studied in detail~ \cite{Yee:2003ge}. As there are
enough supersymmetries, there is no lift of the vacuum energy due
to the mixing by tunnelling.  Similar tunnelling would also
manifest itself for the BPS configurations for a given charge as
there exist a lot of BPS configurations which are separated by
energy barrier. The quantization of the 1/2 BPS configurations
taking into account the tunnelling effect would be interesting.
Still, we expect   many quantum BPS states for a given $N$ and the
BPS angular momentum $J_3$. This leads to a natural index for
counting the BPS states as a function of $N$ and $J_3$.

In the context of Yang-Mills theory on $R\times S^3$, the 1/2 BPS
states with $SO(3)$ angular momentum would correspond to the BPS
chiral conformal operators  with spherical angular momentum. As
all of fuzzy sphere vacua in the matrix theory are gauge
equivalent to the trivial vacuum in the Yang-Mills theory, our 1/2
BPS solution would correspond to a special class of the 1/2 BPS
solutions which has spatial angular momentum $J_3$. Most of the
study of BMN  matrix theory in the Yang-Mills theory context
focuses on the cases of $SO(6)$ angular momentum, and it would be
nice to extend this analysis to our case. (See for example
Ref.~\cite{Plefka:2003nb} for a review.)

The plan of this work is as follows. In Sec.2, we review some
basic aspects of BMN matrix theory and introduce irreducible BPS
configurations. In Sec.3, we perform a perturbative analysis of
the BPS configurations near the abelian solutions and the fuzzy
sphere vacua. In Sec.4, we find some new solutions by writing the
known solution in higher dimensional representations as well as
find some genuinely new solutions. In Sec.5 we conclude with some
remarks.

\section{Plane-Wave Matrix Theory}

In this work, we consider a special class of 1/2 BPS
configurations in BMN matrix theory. BMN matrix theory has
$SO(3)\times SO(6)$ global symmetries and we are here interested
the classical BPS configuration with only $SO(3)$ angular
momentum. Thus we  focus on the part of BMN matrix gauge theory
for three  $N\times N$ hermitian matrices $X_a,a=1,2,3$, whose
Lagrangian is
\be L =  \frac{1}{2} {\rm Tr} (D_0X_a)^2 - U(X) \, ,\ee
where $D_0X_a = \dot{X}_a  -i[A_0,X_a]$, and the potential is
\be U(X) =\frac{1}{2}\, \Tr \left(
\frac{\mu}{3}X_a+\frac{i}{2}\epsilon_{abc}[X_b,X_c]\right)^2 \ee
with $\mu>0$ by a choice of convention.  (Here we have scaled out
the other parameters in the theory for simplicity.) There is a
local time-dependent $U(N)$ gauge symmetry, and any physical
configuration must satisfy the Gauss law constraint,
\be \sum_a  [X_a,D_0X_a]= 0 \, . \label{gauss}\ee
The conserved energy  is
\be H = \frac{1}{2} {\rm Tr}  (D_0 X_a)^2 + U(X)\, . \ee
There is a SO(3) global rotation symmetry of three matrices
$X_1,X_2,X_3$, whose conserved quantities are
\be J_a = \epsilon_{abc} \Tr X_b D_0 X_c \, . \ee

The vacuum configurations are the minima of the potential $U(X)=0$
and satisfy the equation
\be [X_a,X_b] = \frac{i\mu}{3}\epsilon_{abc} X_c\, . \ee
With the scaling $X_a = \frac{\mu}{3}L_a$, $L_a$ form the $SU(2)$
algebra. A partition $(p_1,p_2,,,,,p_K)$ of the number $N$ with
natural numbers $p_k$ such that $\sum_k p_k = N$ characterizes  a
unique gauge equivalent classical vacuum where each $p_k$ denotes
the dimension of the irreducible representation of the $SU(2)$
generators $L_a$.  For example for $N=3$, we have the symmetric
vacuum $(1,1,1)$ where $L_a$ becomes  three trivial 1-dim
representations, the $(2,1)$ vacuum where $L_a$ forms one 2-dim
irreducible representation and one trivial 1-dim one, and the
$(3)$ vacuum where $L_a$ is the 3-dim irreducible representation.
For general $N$, the symmetric phase $L_a=0$ would be denoted as
the $(1,1,...,1)$ vacuum , and the maximally broken vacuum,
  where $L_a$ is the $N$-dim irreducible
representation, would be denoted as the $(N)$ vacuum.  The number
of partitions would be the number of the gauge-equivalent vacua.
As $N$ increases, the number of the partition of $N$ grows very
fast and so is the number of gauge inequivalent vacua.

For a given conserved value, say, $J_3$, the energy can be
reexpressed as
\ba H &=& \frac{1}{2}\left( D_0 X_1 \pm (\frac{\mu}{3}X_2
+i[X_3,X_1])\right)^2 + \left( D_0X_2
\mp(\frac{\mu}{3}X_1+i[X_2,X_3])\right)^2 \nonumber \\
& & +\frac{1}{2} (D_0X_3)^2 + \frac{1}{2}\left(\frac{\mu}{3}X_3
+i[X_1,X_2]\right)^2  \;\; \pm \frac{\mu}{3}J_3 \, . \ea
Thus there is a BPS bound on the energy
\be H \ge \frac{\mu}{3}|J_3|\, ,  \ee
where the Noether charge $J_3$ plays the role of a central term.
(The charge $J_3$ is really the so-called `non-central' term as
the supercharge is not invariant under
it~\cite{Berenstein:2002jq}.)  The bound is saturated for the
so-called BPS configurations which should satisfy  the Gauss law
constraint (\ref{gauss}) and the following BPS equations;
\ba &&  D_0 X_1 = \pm \left(-\frac{\mu}{3}X_2 -i[X_3,X_1]\right),
\;\;\;
D_0X_2 = \pm \left(  \frac{\mu}{3}X_1+i[X_2,X_3] \right) \, , \nonumber \\
&& D_0X_3=0,\;\;\;  \frac{\mu}{3}X_3 +i[X_1,X_2] = 0  \, .
\label{bpseq0}\ea
The upper sign is for $J_3>0$ and the lower sign is for $J_3<0$ .
For BPS configurations, the central charge becomes
\be J_3= \pm \frac{\mu}{3}\;  \Tr ( X_1^2+X_2^2-2 X_3^2) \, . \ee
Thus a BPS configuration of finite energy should satisfy the
inequality
\be \Tr( X_1^2+X_2^2)\ge  2\Tr (X_3^2)\,  .  \ee
As $J_3$ is related to the rotation in the $X_1,X_2$ plane, the
above inequality is consistent with the notion that  that the BPS
solutions are stretched along the 12  plane due to the centrifugal
force.

To find nontrivial solutions, we start with the  gauge choice
$A_0=X_3$ in Eq. (\ref{bpseq0}), and read off the time dependence
of $X_a$. We introduce a complex matrix $W$ and a hermitian matrix
$Z$, both of which are time-independent, and rewrite $X_a$ as
follows;
\be X_1+i X_2=\frac{\mu}{3}e^{\frac{i\mu t}{3}} W\, , \;\;\;
X_1-iX_2=\frac{\mu}{3}e^{-\frac{i\mu t}{3}}\BW\, ,  \;  \;\; X_3
=\frac{\mu}{3}Z\, . \label{td}\ee
The BPS equations and Gauss law in Eqs.(\ref{bpseq0})
and(\ref{gauss}) become $[W,\BW]=2Z$ and
\be  [W,[\BW,Z]]+[\BW,[W,Z]]=4Z \, . \label{mastereq} \ee
These equations are invariant under $SU(N)$ gauge transformations.
In addition, there is an overall $U(1)$ phase rotation of $W$
generated by the charge $J_3$.  A simple solution of the above
equations is
\be W= L_1 + iL_2 = L_+\, , \BW= L_1-iL_2=L_-\, ,\;\; [W,\BW]=
2L_3\, . \label{su2vacuum}\ee
This solution describes the vacuum solution as   the
time-dependent $X_1,X_2,X_3$ in Eq.(\ref{td}) are gauge transforms
of  the vacuum solution $X_a=\mu L_a/3$ by an unitary
transformation $e^{it L_3}$. The vacuum solution has zero central
charge.

We are interested in BPS configurations with nonzero charge $J_3$.
There are many vacua, all of which are separated by some potential
energy barrier. Let us imagine to add a bit of $J_3$ charge at a
given vacuum and find the corresponding BPS configurations in a
given vacuum. We expect many gauge inequivalent BPS configurations
built in a given vacuum  for a given charge $J_3$. Some of these
solutions may be  continuously connected to each other for a given
charge. As we increase the charge, the solutions disconnected from
each other may be get connected as, for example, the
configurations goes over the energy barrier.

To study these BPS configurations, let us start from the abelian
vacuum $(1,1,1,...,1)$ where $X_a=0$, and add some $J_3$ charge.
One can easily find that the purely abelian configurations  where
$W$ is diagonal and traceless satisfy  the BPS equations
(\ref{mastereq}) with $Z=0$. We parameterize the abelian solution
as
\be W= {\rm diag}(\lambda_1,\lambda_2,...,\lambda_N) \, .
\label{abeliansol} \ee
The configuration is completely lying in the 12  plane as $X_3=0$.
The central charge becomes
\be J_3=\frac{\mu^3}{27} \sum_k |\lambda_k|^2 \, .
\label{abelianJ} \ee
There is no upper bound on the value of $J_3$. For a given $J_3$,
the configuration space is parameterized by  $N$ complex numbers.
We will do a small fluctuation analysis of the above abelian
solutions in the next section to find out whether they can be
deformed to nonabelian configurations.

To find nontrivial solutions, let us start from a vacuum
$(p_1,p_2,..,p_K)$ and add a small  $J_3$ charge. There may be
several possible BPS solutions.  The simplest one is the $SU(2)$
type where $X_a$ is a linear combination of $L_a$ whose
representation is characterized by the partition
$(p_1,p_2,...,p_K)$. We start with $W$ being a linear combination
of $L_a$. By a $SU(2)$ gauge transformation, we choose $[W,\BW]$
to be  proportional to $L_3$, which  is diagonal. This in turn
implies that $W$ should be a linear combination of $L_1,L_2$. Then
the solution of the BPS equation  becomes~\cite{Bak:2002rq}
\be  W= c_1 L_+ + c_2 L_- ,\;\; \BW=\bar{c}_1 L_-+\bar{c}_2 L_+
,\;\; Z=(2|c_1|^2-1)L_3  \label{ellip} \, ,\ee
where $|c_1|^2+|c_2|^2=1$. Now we can make a $e^{i\alpha L_3}$
gauge rotation and a  $W\rightarrow e^{i\beta} W$ spatial rotation
to make  both $a,b$ real. Then the solution becomes the well-known
ellipsoid solution
\be \frac{ X^2}{(c_1+c_2)^2} +
\frac{Y^2}{(c_1-c_2)^2}+\frac{Z^2}{(c_1^2-c_2^2)^2} =
\sum_c(L_c)^2 \, . \ee
Since  $c_1^2+c_2^2=1$, the conserved charge $J_3$ becomes
\be J_3= \frac{\mu^3}{27}  \cdot \frac{8}{3}c_1^2(1-c_1^2)\;
\sum_a\Tr(L_a)^2 \, . \ee
Here we have used that $\Tr L_1^2=\Tr L_2^2=\Tr L_3^2=\sum_a
\Tr(L_a)^2/3$. Note that the case where $c_1=1,c_2=0$ or
$c_1=0,c_2=1$ is the vacuum solution and the case where
$c_1^2=c_2^2=1/2$ is the configuration of the collapsed ellipsoid
to an abelian thin line which is rotating on 12 plane. The central
charge $J_3$ takes its maximal value at this abelian limit. Thus
this solution is connected to the abelian solution at the maximal
$J_3$ limit.

For the $N$-dim irreducible representation, $\Tr L_c^2 =
N(N^2-1)/4$. For other representations $(p_1,p_2,...,p_K)$ such
that $\sum_k p_k=N$, $\Tr L_c^2= \sum_k p_k(p_k^2-1)/4 \le
N(N^2-1)/4$. Thus, among the $SU(2)$ type solutions from the
various vacua, the one with with $N$-dim irreducible
representation from the $(N)$ vacuum takes the maximal value,
\be J_{max}= \frac{\mu^3}{27} \frac{N(N^2-1)}{6} \, ,
\label{jmax}\ee
at the abelian limit.

One can easily generalize the above two types of solutions by
putting them together. For any two commuting solutions $W=W_1,
W_2$ of the BPS equations so that $[W_1,W_2]=0, [W_1,\BW_2]=0$,
their sum $W=W_1+W_2$ is also a BPS configuration. Thus in the
vacuum where $L_a$ is not irreducible, we can generalize the above
ellipsoid solution so that each irreducible part has a different
parameter $c_1$ and also one can add abelian solutions which
commute with this generalized nonabelian solution. This leads to a
division of all  BPS configurations into reducible ones and
irreducible ones. The irreducible ones are those which cannot be
expressed as a sum of commuting BPS solutions. Thus all abelian
solutions with $N\ge 2$ are  reducible.

As an example for a reducible BPS configuration, we can consider a
mixed-type BPS configuration built from the  $(2,1)$ vacuum in the
$N=3$ case, which is
\be W=\left(
\begin{array}{ccc}
         c_3 & c_1 & 0 \\
         c_2 &  c_3 & 0 \\
         0 & 0 & -2c_3 \end{array}\right) \, ,
         \ee
where $c_1^2+c_2^2=1$ and $c_3$ is arbitrary. Then $Z=
(c_1^2-1/2)\; {\rm diag}(1,-1,0)$.  When $c_1^2=c_2^2=1/2$, the
above solution can be diagonalized.  When $c_1=1, c_2=c_3=0$, it
becomes the vacuum $(2,1)$. Its angular momentum is
\be J_3= \frac{\mu^3}{27} (6c_3^2+4c_1^2(1-c_1^2) )\, .\ee
Note that $c_3$ can be arbitrary and so there is no bound on $J_3$
for this solution. One can generalize this type of solution
easily. For any nonabelian vacuum where $L_a$ is not irreducible,
there is at least one unbroken abelian $U(1)$ subgroup which
commutes with $L_a$, and so one can add the $J_3$ charge into both
abelian and nonabelian sectors. These type of solutions are
reducible and can be decomposed to a sum of irreducible ones.

We are interested in all BPS configurations. Besides the
ellipsoidal solutions, one may wonder whether there are other
type of nonabelian solutions built on a given nonabelian vacuum
by adding small angular momentum. In the next two sections, we
will add some new understanding on this topic.

\section{Fluctuation Analysis}

For example, the ellipsoidal solution (\ref{ellip}) collapses to
the abelian solution at its maximal $J_3$ value. Thus one suspects
that all nonabelian solutions are connected to abelian solutions
(\ref{abeliansol}). To find out this connection, let us make a
small perturbation of the abelian solution,
\be W= \lambda_i\delta_{ij}  + \epsilon_{ij}  , \;\;\; \BW
=\lambda_i^*\delta_{ij} +\epsilon_{ji}^*  \, , \ee
where $\epsilon_{ii}=0$ for each $i$. Note that a pure gauge
transformation would be $\epsilon_{ij}=
(\lambda_i-\lambda_j)\chi_{ij}$ with antihermitian $\chi_{ij}$.
Then
\be [W,\BW]_{ij} = (\lambda_i-\lambda_j)\epsilon_{ji}^*
-(\lambda_i^*-\lambda_j^*)\epsilon_{ij} \,  . \ee
Then the left-hand-side of the BPS equation (\ref{mastereq})
becomes
\be l.h.s. = 2|\lambda_i-\lambda_j|^2 \Big(
(\lambda_i-\lambda_j)\epsilon_{ji}^*
-(\lambda_i^*-\lambda_j^*)\epsilon_{ij} \Big) \, ,  \ee
which should be $4[W,\BW]$. Thus the BPS equation is satisfied if
for all non-vanishing $\epsilon_{ij}$ which is not a pure gauge
transformation,
\be |\lambda_i-\lambda_j|^2 = 2 \, . \label{crit}\ee
The central charge does not change from the abelian result  to the
first order in perturbation. This analysis shows how any
nonabelian configuration may be connected to the abelian
solutions. This analysis does not tell whether an irreducible
nonabelian solution becomes abelian as $J_3$ increases or
decreases at some critical value, nor tell whether an irreducible
nonabelian solution becomes abelian at once, or piece-wise.

Let us now consider how the above analysis appears in the  the
ellipsoid case (\ref{ellip}). With $N$-dim irreducible $L_a$, the
solution becomes abelian when $c_1=c_2= 1/\sqrt{2}$ as $W=\sqrt{2}
L_1$. The solution $W=\sqrt{2}L_1$ can be diagonalized to be
\be W= \sqrt{2}{\rm diag}(l,l-1,l-2,...,-l+1,-l)\, ,
\label{lines}\ee
where $2l+1=N$. This solution satisfies the criterion
(\ref{crit}). The ellipsoid solution (\ref{ellip}) near this
abelian solution becomes $W= (1-\epsilon^2)L_1 + \epsilon L_2$
which has nonzero $\epsilon_{i,i+1}$.

There are other possible cases. For example, for a given $N$, we
could put $\lambda_i$ on a circle whose center is at the origin.
We require $|\lambda_n-\lambda_{n+k}|^2=2$ for a fixed $k$ and
$\lambda_{N+n}=\lambda_n$. We use the criteria (\ref{crit}) to get
\be \lambda_n=  \frac{1}{\sqrt{2}\sin\frac{\pi k}{N} }
e^{\frac{2i\pi nk}{N}} \label{circles}  \, .\ee
This abelian solution is gauge equivalent to the following one:
\be W=  \frac{1}{\sqrt{2}\sin\frac{\pi k}{N} } S^k_N ,\;\;\; S_N=
\left(\begin{array}{cccccc}
                             0 & 1 & 0 & 0 & ...& 0 \\
                             0& 0& 1 & 0 & ... & 0 \\
                             0 & 0 & 0 & 1 & ... & 0 \\
                             . & . & . & . & ... & . \\
                             0 & 0 & 0 & . & ... & 1 \\
                             1 & 0 & 0 & . & ... & 0
                             \end{array} \right)\, , \label{circles1} \ee
where the $N$-dim shift operator $S_N$ is a unitary matrix with
eigenvalues $e^{2\pi i n /N}$ with  $n=0,1,2,...,N-1$. An
irreducible nonabelian  BPS solution may develop from this abelian
solution with nonzero $\epsilon_{n,n+k}$. Indeed in the next
section, we explore this possibility in detail. The BPS charge
$J_3$ for this critical solution is
\be J_3 = \frac{\mu^3}{27} \frac{N}{2\sin^2\frac{\pi k}{N}} \, .
\ee
For large $N$, the above $J_3$ approaches $
\frac{\mu^3}{27}\frac{N^3}{2\pi^2k^2} $, which is smaller than the
maximal value (\ref{jmax}) for the ellipsoidal case.

For an abelian solution, we draw a line between any pair of
$\lambda_i$'s satisfying the condition~(\ref{crit}). As
$\lambda_i$ lie on a complex plane, a graph made of those lines
can be decomposed to connected graphs. For each connected graph,
there is a potential to develop a nonabelian configuration. Of
course there is no guarantee that nonabelian solutions can
develop. Any part or whole of a nonabelian BPS configuration will
be come such a graph whenever some or all of it becomes abelian.

For the $SU(3)$ case with $\sum_i\lambda_i=0$, let us consider the
case where there points are connected by two lines.  Each of
segment has  length $\sqrt{2}$. They could lie on a straight line
or get bent and form a letter V shape.  They may form an
equi-triangle, or form a sharper tipped V shape, get bent
completely to be a single segment
 where two end points
overlap. All these abelian solutions have a potential to be
nonabelian. For a straight string and equi-triangle cases, the
nonabelian solutions are known. For other bent cases, the
nonabelian extensions, if they exist, would be an interesting
possibility. Now one can see easily that the central charge
(\ref{abelianJ}) for these configuration takes the maximal value
for straight line. Similarly, we conjecture that the central
charge $(\ref{abelianJ})$ of connected graphs of N points take the
maximal value for the N points lying on a straight line
(\ref{lines}).

Now let us change our focus to the BPS configurations built on
nonabelian vacua. We start from a nonabelian vacuum where $L_a$ is
nontrivial and add a small amount of $J_3$. A BPS configuration
close to the vacuum can be approached by perturbation analysis. Of
course there is an ellipsoidal solution (\ref{ellip}) near each
vacuum.  We deform the vacuum solution by a small deformation,
\be W= L_+ +\delta W \, .\ee
We first focus on the $(N)$ vacuum case and expand the matrix
$\delta W$ as the sum of irreducible representations of $L_a$. For
example, the (3) vacuum in the $N=3$ theory, $SU(3)$ generators
belongs to 5-dim and 3-dim representations. The $N$-dim matrices
$T^l_m $, which belong to the $l$ representation, satisfy the
commutation relations,
\be [L_\pm, T^l_m]= \sqrt{l(l+1)-m(m\pm1)} T^l_{m\pm 1} ,\;\;
[L_0,T^l_m]=m T^l_m \, , \ee
which is consistent if
$ T_m^{l\dagger} = (-1)^m T^l_{-m}\, . $
The normalization is $\Tr T^l_m T^{l'}_n \sim
\delta_{ll'}\delta_{m+n,0}$. For example $T^1_1=L_+/\sqrt{2}$, $
T^1_0=-L_0$, $T^1_{-1}=-L_-/\sqrt{2}$ for the obvious $l=1$
representation. In this basis, the fluctuated $W$ becomes
\begin{equation}
  W= L_+ + C^l_m T^l_m , \,  {\bar W} = L_- + (-1)^m \bar{C}^l_{-m}
T^l_m \end{equation}
with complex coefficients $C^l_m$. We remove the gauge degrees of
freedom by diagonalizing  $[W,\BW]=2Z$, which leads to the
following relations among the coefficients,
\be (-1)^m\bar{C}^l_{-m+1} A_{m-1}+C^l_{m+1} B_{m+1}= 0 , \;\;
m\neq 0 , \label{coef1} \ee
where $A_m= \sqrt{l(l+1)-m(m+1)}$ and $B_m=\sqrt{l(l+1)-m(m-1)}$.
This condition implies
\be C^l_{-l}= {\rm arbitrary}, \;\; C^l_{-l+1}=0 \, ,\ee
and
\be [W,\BW]= 2L_3-\sqrt{l(l+1)}(\bar{C}^l_1+C^l_1)T^l_0 \, .
\label{rhsz} \ee
The left-hand side of the BPS equation (\ref{mastereq}) becomes
\ba &&  l.h.s. =
8L_3-2(l^2+l+2)\sqrt{l(l+1)}(C^l_1+\bar{C}^l_1)T^l_0 \nonumber
\\ && \!\!\!\!\!\! -2\sum_{m\neq 0}\Big[
(-m+2)(-1)^m\bar{C}^l_{-m+1}A_{m-1}+(m+2)C^l_{m+1}B_{m+1}\Big]T^l_m
\, . \ea
Equating the above expression with four times the expression in
Eq.~(\ref{rhsz}), we get  that for $l\neq 1$ $C^l_m$ vanishes for
all $m$ except $m=-l$, and for $l= 1$, $C^1_1,C^1_{-1}$ can be
arbitrary and $C^1_0$ vanishes.

Thus from the vacuum $(N)$, the BPS configurations near the vacuum
$X_a=L_a$ should be given as
\be W= L_a + \sum_{l\neq 1} C^l_{-l}T^l_{-l} + C^l_1 T^1_1
+C^1_{-1}T^1_{-1}  \, ,  \ee
where one can sum over all $l$ which appear when the $SU(N)$
generators are split into irreducible representation of
irreducible $SU(2)$ algebra.   Note that by  gauge transformation
and  $U(1)$ rotation one can make $C^1_1$ and $C^1_{-1}$ to be
real. The above analysis shows that the ellipsoid solution $W=c_1
L_+ + c_2 L_-$ grows out of the above fluctuation with nonzero
$C^1_1, C^1_{-1}$.

Let us turn our attention to other vacua, for example,  the vacuum
$(p_1,p_2,...,p_K)$ with $\sum_k p_k = N$,  in which case
$L_a=L_a^1\oplus L_a^2\oplus...$ with each $L_a^k$ being $p_k$-dim
irreducible representation. Our linear fluctuation analysis can be
extended directly. The fluctuation is replaced by
$C^{l_1l_2..}_{m_1m_2..}T^{l_1l_2..}_{m_1m_2..}$ where $(l_k,m_k)$
indicates the representation of the k-th irreducible part of
$L_a^k$.  A coefficient can be nonzero if  for all $k$ either
$m_k=-l_k$ for or $l_k=1, m_k=1$. This is how a BPS solution will
develop when a small amount of $J_3$ charge is added to a
nonabelian vacuum.

\section{Some Exact BPS Solutions}

Let us try to find some (new) exact solutions of the BPS equation
(\ref{mastereq}), that is, of
\be \frac{1}{2}[W^2,\BW^2]=(W\BW)^2-(\BW W)^2-[W,\BW]\, .
\label{quart0} \ee
One could, for example,  assume  two polynomial equations
\ba && \frac{1}{2}W^2\BW^2=(W\BW)^2-W\BW +
f(W\BW, \BW W) \, , \label{quarta} \\
&& \frac{1}{2}\BW^2W^2 = (\BW W)^2-\BW W + f(W\BW,\BW W) \, ,
\label{quartb} \ea
where $f(W\BW,\BW W) $ is an arbitrary polynomial of $W\BW$ and
$\BW W$. Depending on the function $f$, the above equations
(\ref{quarta}) and (\ref{quartb}) could be independent and
over-constraining. Our favorite example is
\be f= \alpha \BW W^2\BW+ \beta W\BW^2 W+ q \label{ff} \ee
with real constants $\alpha, \beta, q$. Requiring the hermicity of
$f$ leads to either $\alpha=\beta$ or $[W\BW,\BW W]=0$.

The above polynomial equations are a slightly more general
non-singular variant of the commutation relation
\be [Z,W]=W-\frac{q}{\BW} \label{toruseq}, \ee
that were used in \cite{Mikhailov:2002wx,Bak:2005jh} to solve the
BPS equation (\ref{mastereq}). By multiplying $\BW$ on the above
equation from the right or left, we get  Eqs.~(\ref{quarta}) and
(\ref{quartb}) with $f$ in Eq.~(\ref{ff}) with $\alpha=-1/2$ and
$\beta=0$. Thus we get $[W\BW,\BW W]=0$ for the BPS configurations
to satisfying (\ref{toruseq}). As noted in
Ref.~\cite{Mikhailov:2002wx}, the BPS equations (\ref{mastereq})
can be solved partially if there exists an analytic function $F$
such that
\be [Z, W]= W + F(\BW) \, . \ee
Assuming that  $W$ is invertible and that there is a $U(1)$ phase
rotation symmetry in the above equation, one gets (\ref{toruseq})
in general.

We consider that the  $[W\BW,\BW W]=$ case in
Eqs.~(\ref{quart0},\ref{quarta}, \ref{quartb}) is very interesting
and may leads to the new type of solutions. But we will not pursue
this direction in this work. We will focus here on the simpler
case (\ref{toruseq}) which has a $U(1)$ symmetry.  When $q=0$,
$W,Z$ satisfy the $SU(2)$ algebra and so the above equation
degenerates to the vacuum equation. As it has a $U(1)$ symmetry
and $W\neq 0$ for $q\neq 0$, we call the BPS configurations
satisfying the above equation as "of toroidal type". Multiplying
by $\BW$ and taking the trace, we get
\be qN= \Tr(W\BW -2Z^2) \, . \ee
The central charge of any configuration of such ansatz would be
\be J_3 = \frac{\mu^3}{27}\Tr(W\BW -2Z^2) = \frac{qN \mu^3}{27} \,
 \ee
which should be positive. Thus we restrict to  $q>0$.

We try to solve the above torus-type equation (\ref{toruseq}) with
the ansatz
\be  W= \left( \begin{array}{ccccc}
   0 & w_1 & 0 &... & 0 \\
   0 & 0 & w_2 & ...  & 0 \\
   . & .& . & .& .\\
   0 & ..&...& 0 & w_{N-1} \\
   w_N & 0 & ..& ..& 0 \end{array} \right) \, , \label{tori}  \ee
so that $W_{ij} = w_i \delta_{i+1,j},\; {\rm mod} \;N$. Here we
find some additional new solutions and some new generalizations.
Both $W\BW$ and $\BW W$ are diagonal and
\be (W\BW)_{ii}= |w_i|^2 \equiv r_i,\;\; (\BW
W)_{ii}=|w_{i-1}|^2\equiv  r_{i-1} \, . \ee
Consistency then requires the of alpha and beta to be -1/2, and
\be r_i \left(\frac{r_{i-1}+r_{i+1}}{2}\right) = r_i^2-r_i + q
\,  . \ee
If $W$ is invertible so that all $r_i\neq 0$, we divide the above
equation by $r_i$ to get
\be r_i + \frac{q}{r_i} = \frac{r_{i-1}+r_{i+1}}{2}+1 \,
.\label{reqs} \ee
When we sum over $i$, we get the constraint
\be \frac{1}{N}\sum_i \frac{1}{r_i} =\frac{1}{q}   \, , \ee
which implies that the average of the inverse $1/r_i$ is $1/q$.

The obvious solution would be $r_i=q$ independent of $i$, which is
abelian as shown in Eqs.(\ref{circles}) and (\ref{circles1}) with
$\lambda_k = \sqrt{q}e^{i2\pi k /N}$. The analysis in the previous
section implies that there may be nonabelian solutions
$\{r_1,r_2,...,r_N\}$ near this constant solution if $q$ is close
to the abelian value $q=1/ (2\sin^2(\pi k/N))$.

Let us find some explicit solution by starting with $N=2$ case.
The explicit solution for the equation (\ref{reqs}) for $N=2$ case
is
\be r_1= \frac{1\pm \sqrt{1-2q}}{2},\;\; r_2= \frac{1\mp
\sqrt{1-2q}}{2}    \, . \ee
This is the ellipsoidal solution (\ref{ellip}) with $N=2$ as we
can identify $c_1^2=r_1, c_2^2=r_2$. There are  two obvious
generalization of this $N=2$ solution. First is somewhat trivial
as one finds the solution for all even $N=2K$ with the periodic
condition $r_{k+2}= r_k$ for all $k=1,2,..,K$. Another one is the
ellipsoidal solution $W=c_1 L_++ c_2 L_-$ as in Eq. (\ref{ellip})
where  $L_a$ are $N$-dim representation of $SU(2)$. If we consider
an irreducible $N$-dim representation of $SU(2)$, we know that
this generalized BPS configuration is also irreducible. This
solution is not new nor does not belong to the toroidal type
(\ref{tori}).

For $N=3$, there is no solution with all different $r_1,r_2,r_3$.
The type of solution that appeared in Ref.~\cite{Bak:2005jh} was
\be r_1=r_2=1\pm \sqrt{1-\frac{4q}{3}}, \;\; r_2=
\frac{1}{2}\left(1\mp \sqrt{1-\frac{4q}{3}}\right) \, , \ee
where $0\le q \le 3/4$. As in $N=2$ case, we have two
generalization of this solution. The first one is the periodic
extension for  all $N=3K$ with $r_{k+3}=r_k$. Another one is to
rewrite the above solution in terms of $SU(3)$ generators and to
generalize the solution to the $N$-dim presentation of the $SU(3)$
Lie-algebra. The above solution is rewritten as
\be W = c_1 L_+ + c_2 P_-   \, ,  \ee
where real $c_1,c_2$ satisfy  $c_1^2+c_2^2 =1$ with $c_1^2=r_1/2,
c_2^2= r_3$, and
\be L_+ =\left(\begin{array}{ccc} 0 & \sqrt{2} & 0 \\
0 & 0& \sqrt{2} \\
0& 0& 0 \end{array}\right),\;\;\; P_-= \left(\begin{array}{ccc} 0&
0& 0 \\ 0& 0& 0 \\ 1 & 0 & 0 \end{array}\right) \, . \ee

With the definition $L_-=L_+^\dagger$ and $P_+=P_-^\dagger$, we
note that $[P_+,P_-]=-L_3$ with $L_3= {\rm diag}(1,0,-1)$ and
$[L_3/2, P_+]=P_+$. Thus  $P_+, P_-, L_3/2$ are another $SU(2)$
generators in $SU(3)$ lie algebra. Thus,
\be Z= (c_1^2-\frac{c_2^2}{2})L_3 =\frac{3c_1^2-1}{2} L_3\, . \ee
The conserved central charge becomes
\be J=\frac{\mu^3}{27} \frac{3c_1^2(1-c_1^2)}{2}  \sum_a  \Tr
(L_a)^2   \, , \ee
where we used the isotropy to show
$\Tr(L_1P_1-L_2P_2)=\Tr(L_1P_2+L_2P_1)=0$ and $\Tr L_i^2 = 4\Tr
P_i^2 = \sum_a L_a^2/3$. We see that when $N=3$ this solution
interpolates between two vacua, $(3)$ at $c_1=1$ and $(2,1)$ at
$c_1=0$, via the abelian BPS  solution at $c_1=1/\sqrt{3}$. Note
that $c_1=1/\sqrt{3}$, $r_1=r_2=r_3=2/3$ and $q=2/3$, and so the
solution becomes abelian. Our solution can be regarded as a
nonabelian BPS solution growing out of this abelian solution. The
maximum value of the central charge for this type of solution
appears at $c_1^2=1/2$ which is not the abelian case $c_1^2=1/3$.
Thus one can see that the abelian solution can be developed to the
nonabelian ones  by either adding or subtracting the central
charge. Still the maximum value of the central charge of this type
is smaller than that of the abelian limit of the ellipsoidal type
(\ref{ellip}).

Now we generalize $L_+$ and $P_-$  to an arbitrary $N$-dim
representation of $SU(3)$.  For example the symmetric $K$ product
of ${\bf 3}$-dim representation of $SU(3)$ would be
$N=(K+1)(K+2)/2$ dimensional irreducible representation of
$SU(3)$. The adjoint representation is $8$-dimensional. For
general integer $N$, the representation would not be irreducible.
Even when we have an  $N$-dim irreducible representation of
$SU(3)$, we do not have a maximal $N$-dim representation of
$SU(2)$ generator $L_a$. For $N=3$ solution, $L_+$ is for 3-dim
representation of $SU(2)$ in $SU(3)$ and $P_a$ is for 2-dim
representation of $SU(2)$ in $SU(3)$. The 6-dim irreducible
representation  of $SU(3)$, for example, belongs to the reducible
$5+1$-dim representations of $L_a$ and the reducible $3+2+1$-dim
representations of $P_a$. Thus our BPS solution interpolates two
vacua,  $(5,1)$ at $c_1=1$ and $(3,2,1)$ at $c_1=0$, via an
abelian BPS solution at $c_1=1/\sqrt{3}$.  By going to higher
  $N$-dim representation, our fuzzy geometry becomes dense and could goes
  to continuum once an appropriate scaling is taken. This process
  could fix the genus uniquely. Of course we expect the present
  solution has torus topology in the continuum limit.

Now for $N=4$, there are two types of solutions given in
Ref.~\cite{Bak:2005jh}. There is no solution where all $r_i$ are
different. First one is the case where $r_1=r_2= 1+\sqrt{1-q}$ and
$r_3=r_4=1-\sqrt{1-q}$ where $0\le q \le 1$. This solution can be
generalized to $N=4K$ cases with $r_{a+4}=r_a$. Also similar to
$SU(2)$ and $SU(3)$ cases, one can reexpress this solution as
\be W= c_1 L_+ + c_2  P_-    \, , \ee
where real $c_1, c_2$ satisfy $c_1^2+c_2^2=1$, and
\be L_+=\left(\begin{array}{cccc} 0 & \sqrt{2} & 0 & 0 \\
    0& 0& \sqrt{2} & 0 \\ 0 & 0& 0& 0 \\ 0 & 0& 0& 0\end{array}
    \right),\;\; P_-=\left(\begin{array}{cccc} 0 & 0 & 0 & 0 \\
    0& 0& 0 & 0 \\ 0 & 0& 0& \sqrt{2} \\ \sqrt{2}  & 0& 0& 0\end{array}
    \right)\, .\ee
Note that $[L_+,L_-]=[P_+,P_-]= 2 L_3$ where $L_3={\rm
diag}(1,0,-1,0)$. The commutation relation leads to
\be Z= (2c_1^2-1) L_3   \, . \ee
The conserved central charge becomes
\be J_3= \frac{\mu^3}{27} \frac{8}{3}c_1^2(1-c_1^2)\Tr L_a^2   \,
. \ee
This solution interpolates the vacuum (3,1) at $c_1=0,1$ to itself
via an abelian BPS solution at $c_1=1/\sqrt{2}$. Again, we can
consider $N$-dim representation of $SU(4)$ and generalize the
above solution to such a space. For example, the adjoint
representation would be $15$-dimensional. The symmetric product of
two 4-dim representation would be 10-dim, and the anti-symmetric
product of two 4-dim representation would be 6-dim.

The second type of solution with $N=4$ is
\ba && r_1=r_3=\frac{1}{2}(3\pm\sqrt{9-8q}),\;  r_2=
\frac{1}{2}(r_1+1\pm \sqrt(r_1+1)2-4q) , \nonumber \\ && r_4=
\frac{1}{2}(r_1+1\mp \sqrt(r_1+1)2-4q)   \, .  \ea
Again this solution can be generalized to $N=4K$ with
$r_{a+4}=r_a$.  We reexpress the above solution as
\be W= c_1L_+ +c_2 M_+ + c_3 P_-    \, , \ee
where
\be L_+ = \left(\begin{array}{cccc}
            0& 1  & 0 & 0 \\
            0& 0& 0  & 0  \\
            0& 0& 0&  1 \\
            0 & 0& 0& 0 \end{array}\right)\, ,\;
M_+= \left(\begin{array}{cccc}
            0& 0 & 0 & 0 \\
            0& 0& 1& 0  \\
            0& 0&  0& 0 \\
            0 & 0& 0& 0 \end{array}\right)  \, ,\;
P_-= \left(\begin{array}{cccc}
            0& 0 & 0 & 0 \\
            0& 0& 0 & 0  \\
            0& 0& 0& 0 \\
            1 & 0& 0& 0 \end{array}\right)\,   . \;\ee
Here the coefficients $c_i$ are given by the relations,
$c_1^2=r_1, c_2^2=r_2, c_3^2=r_2$, and satisfy the constraints
$c_1^2+1=c_2^2+c_3^2$ and $c_1^2(3-c_1^2)=2c_2^2 c_3^2$. The range
of $c_1$ is   $0\le c_1^2 \le 1/3$ or $1\le c_1^ 2\le 3$. The
commtation of $[W,\BW]$ leads to
\be Z =  c_1^2L_3+c_2^2 M_3-c_3^2 P_3 \, ,  \ee
where
\be L_3= \frac{1}{2} {\rm diag}(1,-1,1,-1),\;\; M_3= \frac{1}{2}
{\rm diag}(0,1,-1,0), \;\;  P_3= \frac{1}{2} {\rm
diag}(1,0,0,-1)\, . \ee
The central charge becomes
\be J = \frac{\mu^3}{27} \frac{4}{3} c_1^2(3-c_1^2)\Tr M_a^2 \,  ,
\ee
where we have used $L_3= -M_3+P_3$ and $\Tr L_a^2=2 \Tr M_a^2=2\Tr
P_a^2$. When  $c_1^2=0$, $(c_2^2,c_3^2)=(0,1) $ or $(1,0)$, the
solution becomes the $(2,1,1)$ vacuum. When $c_1^2=3$,
$(c_2^2,c_3^2)=(0,4),$ or $(4,0)$ the solution becomes the $(4)$
vacuum. When $c_1^2=1/3$, $c_2^2=c_3^2=2/3$, it is a nonabelian
BPS solution obtained from the N=2 case with the generalization
$r_{i+2}=r_i$. When $c_1^2=1$, $c_2^2=c_3^2=1$ and so that $Z=0$
and so the solution becomes the abelian BPS solution.

For N=5, one can consider type of solution $r_1=r_4, r_2=r_3, r_5$
type of solutions which is invariant under the reflection of a
pentagon. One gets a 7th order equation which is hard to solve.
However, one can in principle generalize this solution to the
operator equations,
\be W= c_1 L_+  + c_2  M_+ + c_3  P_- \, , \ee
where
\be L_+ = \left(\begin{array} {ccccc}
            0 & 0 & 0& 0& 0 \\
            0 & 0 & \sqrt{2} & 0 & 0 \\
            0 & 0& 0& \sqrt{2} & 0 \\
            0 & 0& 0& 0& 0 \\
            0 & 0& 0& 0 & 0
            \end{array} \right) ,\;\;
M_+ = \left(\begin{array} {ccccc}
            0 & 1 & 0& 0& 0 \\
            0 & 0 & 0 & 0 & 0 \\
            0 & 0& 0& 0 & 0 \\
            0 & 0 & 0 & 0 & 1 \\
            0 & 0& 0& 0& 0
            \end{array}  \right) ,\;\;
P_- = \left(\begin{array} {ccccc}
            0 & 0 & 0& 0& 0 \\
            0 & 0 & 0 & 0 & 0 \\
            0 & 0& 0&  0 & 0 \\
            0 & 0& 0& 0& 0 \\
            1 & 0 & 0 & 0 & 0
            \end{array} \right) \, .
\ee
The coefficients $c_i$ satisfy the relations, $(c_2^2+1)c_3^2
-c_3^4= (2+c_2^2)c_1^2-2c_1^4$ and $2(c_2^2+c_1^2)c_3^2-2c_3^4
=(2+2c_1^2+c_3^2)c_2^2-2c_2^4$.  These equations are easer to
solve, say for a given value of $c_3$.

The equation for $r_i$ becomes more and more involved as $N$
increases. We find a  new exact solution with $N=6$,  which does
not appear in Ref.~\cite{Bak:2005jh} , such that
\ba && r_3=r_6= 2\pm \sqrt{4-2q}  \,  , \nonumber  \\
&& r_1=r_2= (1+ r_3/2)\pm \sqrt{(1+r_3/2)2-2q} \, , \nonumber \\
&& r_4=r_5= (1+r_3/2)\mp \sqrt{(1+r_3/2)2-2q} \, . \ea
In terms of matrices, the solution can be written as
\be W= c_1 L_+ + c_2 M_+ + c_3 P_+ \ ,  , \ee
where
  \ba && L_+ = \left(\begin{array} {cccccc}
            0 & \sqrt{2} & 0& 0& 0 &0 \\
            0 & 0 & \sqrt{2} & 0 & 0 &0\\
            0 & 0& 0& 0 & 0 &0\\
            0 & 0& 0& 0& 0 &0 \\
            0 & 0& 0& 0& 0& 0  \\
            0 & 0& 0& 0& 0& 0
            \end{array}\right)\,  ,\;\;
  M_+ = \left(\begin{array} {cccccc}
            0 & 0& 0& 0& 0 &0 \\
            0 & 0 & 0& 0 & 0 &0\\
            0 & 0& 0& 0 & 0 &0\\
            0 & 0& 0& 0& \sqrt{2} &0 \\
            0 & 0& 0& 0& 0& \sqrt{2}  \\
            0 & 0& 0& 0& 0& 0
            \end{array}\right) \, ,\;\; \nonumber \\
  &&
  P_+ = \left(\begin{array} {cccccc}
            0 & 0& 0& 0& 0 &0 \\
            0 & 0 & 0& 0 & 0 &0\\
            0 & 0& 0 & 1 & 0 &0\\
            0 & 0& 0& 0& 0  &0 \\
            0 & 0& 0& 0& 0& 0   \\
            1 & 0& 0& 0& 0& 0
            \end{array}\right) \, . \;\;
\ea
The coefficients are given as $2c_1^2=r_1=r_2$, $2c_2^2=r_4=r_5$
and $c_3^2=r_3=r_6$, and so satisfy the constraints,
$c_1^2+c_2^2=1+c_3^2/2$, and $4c_1^2 c_2^2 =c_3^2(4-c_3^2)$. The
range is given by $0\le c_3^2\le 2/5$ or $2\le c_3^2\le 4$. The
commutation relation $[W,\BW]$ leads to
\be Z= c_1^2 L_3 + c_2^2 M_3 +c_3^2 P_3 \, , \ee
where
\ba && L_3 = {\rm diag}(1,0,-1,0,0,0)\, ,  \nonumber \\
&& M_3 ={\rm diag}(0,0,0,1,0,-1)\, ,  \nonumber \\
&& P_3 = \frac{1}{2} {\rm diag}(-1,0,1,-1,0,1)\, .  \ea
Note that $P_3= -(L_3+M_3)/2$.  The central charge becomes
\be J_3 = \frac{\mu^3}{27}\frac{2}{3} \Big(
\big(c_1^2+\frac{c_3^2}{4}-(c_1^2-\frac{c_3^2}{2})^2 \big)\Tr
L_a^2 + \big(c_2^2+\frac{c_3^2}{4} -(c_2^2-\frac{c_3^2}{2})^2\big
) \Tr M_a^2\Big) \, .\ee
This solution interpolates the vacuum (3,1,1,1) at
$c_1^2=1,c_2^2=0,c_3^2=0$ or $c_1^2=0,c_2^2=1,c_3^2=0$ and the
vacuum (5,1) at $c_1^2=3,c_2^2=0,c_3^2=4$ or
$c_1^2=0,c_2^2=3,c_3^2=4$, via the abelian solution at
$c_1^2=c_2^2=1,c_3^2=2$. Again this solutions can be generalized
to arbitrary dimension by considering $N$-dim representation of
the $SU(6)$ generators

\section{Conclusion}

In this work, we investigated the 1/2 BPS configurations with
$SO(3)$ angular momentum in the BMN matrix theory. From the
abelian BPS configurations, we have seen how nonabelian field
configurations can emerge and made a conjecture on the exact value
of the maximum angular momentum $J_3$ as a function of $N$ for
irreducible nonabelian configurations. From the fluctuation
analysis of the BPS configurations around the nonabelian vacuum we
learned how nontrivial solutions can emerge  from a given fuzzy
sphere. Finally we found some exact 1/2 BPS configurations which
are new because of the further interpretation of the already known
ones or some new type of solutions.

The general solutions of the 1/2 BPS equations have been found in
the continuum limit in Ref.~\cite{Bak:2005jh}. The BPS solutions
can be characterized by the Riemann surface with arbitrary number
of genus and spikes. For finite $N$, it is more difficult to
assign the genus to the fuzzy object. For an irreducible
nonabelian BPS configuration in a given $N$, one can take a unique
large $N$  and continuing limit by taking the higher dimensional
irreducible representation of the same configuration. This may let
us assign a unique genus number to  a given irreducible fuzzy
object. Our exact BPS solutions are fuzzy tori type of  genus 1. A
variation of our approach may leads to more involved fuzzy Riemann
surfaces. This remains to be seen.

The 1/2 BPS deformations of the maximally supersymmetric AdS
geometries in M theory have  been studied
extensively~\cite{Lin:2004nb}. Our 1/2 BPS configurations in the
BMN matrix would correspond to the 1/4 BPS deformation of the AdS
geometries. A direct supersymmetric geometry for BMN matrix theory
has been studied ~\cite{Lin:2004kw}. Other classes of 1/4 BPS
deformations of the AdS geometry in M theory have  been also
explored recently ~\cite{Gauntlett:2004zh,Kim:2007hv}. It would be
interesting to find the geometric counterpart for  the BPS fuzzy
Riemann surfaces studied here.

\noindent{\bf Acknowledgement}

We thank Seok Kim for useful discussions, and Choonkyu Lee for
kind hospitality. This work is supported in part by the Swedish
Research Council (JH), the KOSEF SRC Program through CQUeST at
Sogang University (KML), KRF Grant No. KRF-2005-070-C00030 (KML),
and the KRF National Scholar program (KML).

\end{document}